\documentclass[twocolumn,amssymb,amssymb,aps,pra,showpacs]{revtex4}
\usepackage{float}
\usepackage{epsfig}
\usepackage{amsfonts}
\usepackage{amsmath}
\usepackage{amsthm}
\begin{document}

\title{Potential flows in the Reissner-Nordstr\"om-(anti) de Sitter 
metric: numerical results}

\author{Gustavo C. Colvero} \email[e-mail: 
]{gustavo.colvero@ufabc.edu.br}

\author{Maximiliano Ujevic}\email[e-mail: ]{mujevic@ufabc.edu.br}

\affiliation{Centro de Ci\^encias Naturais e Humanas, Universidade 
Federal do ABC, 09210-170, Santo Andr\'e, S\~ao Paulo, Brasil}

\begin{abstract}

Numerical solutions for the integral curves of the velocity field 
(streamlines), the density contours, and the accretion rate of a 
steady-state flow of an ideal fluid with $p=K n^\gamma$ equation of 
state are presented. The streamlines and velocity fields associated with 
a black hole and a rigid sphere in a Reissner-Nordstr\"om-(anti) de 
Sitter spacetimes are studied in some detail. For each case the fluid 
density is studied using contour lines. For $\gamma \neq 2$, we found 
that the studied properties of the fluid are more sensitive to 
variations of the electric charge and the cosmological constant. Also, 
the accretion rate was found to increase or decrease when the 
cosmological constant increases or decreases respectively.

\end{abstract}

\pacs{47.15.Hg, 04.25.Dm, 47.75.+f, 97.60.Lf}

\maketitle

\section{Introduction}

Most of the articles about potential flows in this area deal with the 
important case of fluid motion evolving in the spacetime associated with 
compact stars and black holes. The implementation of new background 
metrics brings some new challenges. First, metrics other than 
Schwarzschild and Kerr are not so well studied, sometimes a complete 
understanding of the physical meaning of the metric is missing.  Also, 
the solutions of the fluid equations in a non trivial metric may be 
quite involved. In particular, the search for significant boundary 
values (or initial conditions) presents a non trivial problem. A simple 
and paradigmatic case of a flow is the stationary-zero-vorticity flow of 
a fluid with adiabatic stiff equation of state. In this case, for 
relativistic flows, the fluid equations admit analytical solutions for 
some particular metrics \cite{sha,pet:sha,bab:che}. These solutions are 
used as test-beds for testing almost all the numerical hydrodynamic 
codes in the subject. Other potential flows in a nonstationary 
background and different equation of state had been studied, see for 
instance \cite{abr:sha,uje:let1,uje:let2}. Also, the solutions for 
potential flows permit to test new optimized codes in resolving 
nonlinear hyperbolic systems of conservation laws \cite{fon:mar}, see 
for a representative sample of numerical schemes \cite{fon}.

In previous articles \cite{sha,pet:sha,uje:let1,bab:che} the authors 
used for the fluid the stiff equation of state, i.e., a polytropic, 
$p=Kn^\gamma$ with $\gamma=2$.  This stiff equation of state over 
simplifies the partial differential equation for the velocity potential 
field $\Phi$ (it becomes linear). In this case the sound velocity in the 
fluid is equal to the speed of light. Therefore the stiff equation of 
state represents a limit situation not easily encountered in usual 
physical situations. Some articles \cite{abr:sha,uje:let2} studied the 
more realistic situation, a fluid with a polytropic equation of state 
with $ 1<\gamma<2 $ in the presence of either a rigid sphere or a black 
hole.  The sound velocity in this case is less than the speed of light 
and the partial differential equation for $\Phi$ turns to be nonlinear 
and needs to be solved numerically. In this work we present a nontrivial 
extension of the investigations about potential flows by studying the 
streamlines and baryon density contours for a fluid with polytropic 
equation of state with $1<\gamma<2$ in the presence of either a rigid 
sphere or a black hole in a Reissner-Nordstr\"om-(anti) de Sitter 
background. We consider also the accretion rate of particles into the 
black hole and its dependence with $\gamma$, the electric charge and the 
cosmological constant. We assume that the fluid is a test fluid, i.e., 
the metric does not evolve and it is given {\em a priori}. The state 
equation and the idea of a rigid star (rigid sphere) we use are 
idealized. However, they bring important results about the behavior of 
the fluid, and the difficulties involved in this kind of scenarios.

This work is divided as follows. In Sec. II we present the basic 
equations that describe potential flows and the nonlinear equation for 
the velocity potential for a polytropic equation of state of the form 
$p=K n^\gamma$. In Sec. III we present the numerical method used to 
solve the nonlinear partial differential equation presented in Sec. II. 
In Sec. IV we show the numerical results for the density contours, the 
streamlines and the accretion rate for the black hole and rigid sphere 
cases for different values of the parameters $\Lambda$, $M$, $Q$ and 
$\gamma$. Finally, in Sec. V we summarize our results.

\section{Basic Equations}

We start from the energy-momentum tensor for an ideal fluid, $T_{\mu\nu} 
= (p + \rho) U_\mu U_\nu + p g_{\mu\nu}$, where $p$ is the pressure, 
$U^\mu$ represents the fluid four-velocity, $\rho =\rho_0 + \epsilon$ is 
the total energy density, $\rho_0$ is the rest mass energy density, and 
$\epsilon$ is the internal energy density. Our conventions are $G=c=1$, 
metric with signature +2. Partial and covariant derivatives are denoted 
by commas and semicolons, respectively. The conservation equations, 
$T^{\mu\nu}_{;\nu} = 0$, for this kind of fluid reduce to
\begin{equation}
(\rho + p) U^\mu_{;\mu} + \rho_{,\mu} U^\mu = 0, \label{motion1}
\end{equation}

\noindent and
\begin{equation}
(\rho + p)U^\nu U_{\mu;\nu} + p_{,\mu} + p_{,\nu} U^\nu U_\mu = 0, 
\label{motion2}
\end{equation}

\noindent which are, respectively, the `mass' conservation law and the 
Euler equation. For isentropic flows we have $(\sigma/n)_{;\mu}=0$, 
where $\sigma$ is the entropy per unit volume and $n$ is the baryon 
number density. In this case the equations of motion (\ref{motion2}) 
take the form \cite{lan:lif}
\begin{equation}
U^\nu \omega_{\mu\nu} =0,
\end{equation}

\noindent where $\omega_{\mu\nu}$ is the relativistic vorticity tensor 
defined as
\begin{equation}
\omega_{\mu\nu} = \left[ \left( \frac{\rho + p}{n} \right) U_\mu 
\right]_{;\nu} - \left[ \left( \frac{\rho + p}{n} \right) U_\nu 
\right]_{;\mu}.
\end{equation}

\noindent The potential flow solution of this equation 
($\omega_{\mu\nu}=0$) is
\begin{equation}
h U_\alpha = \Phi_{,\alpha}, \label{potflow}
\end{equation}

\noindent where $h=(\rho +p)/n$ is the enthalpy per baryon. From 
Eq.~(\ref{potflow}) and the baryon number density conservation equation,
\begin{equation}
(n U^\mu)_{;\mu}=0, \label{consern}
\end{equation}

\noindent we obtain the differential equation for the velocity 
potential,
\begin{equation}
\Box \Phi + [\ln \left( \frac{n}{h} \right) ]_{,\alpha}
\Phi^{,\alpha}=0, \label{diffeqn}
\end{equation}

\noindent where, $\Box \Phi = [ \sqrt{-g} g^{\mu \nu} \Phi_{,\mu} 
]_{,\nu} / \sqrt{-g}$. This equation is in general nonlinear and depends 
on the fluid equation of state. If we consider a polytropic equation of 
state $p= K n^{\gamma}$ and use the first law of thermodynamics for 
isentropic flows, ${\rm d} (\rho/n) + p {\rm d} (1/n) = T {\rm d} s$, 
($s$ is the entropy per baryon and $T$ the temperature), we find
\begin{equation}
\rho = \rho_0 + \frac{Kn^\gamma}{\gamma-1}.
\end{equation}

Now, assuming that the flow is relativistic we can neglect the rest mass 
energy density $\rho_0$ with respect to the internal energy density $K 
n^\gamma/(\gamma-1)$, i.e., the flow satisfies a barotropic equation of 
state, $p=(\gamma -1)\rho$. In this case, the baryon number density can 
be written in terms of the enthalpy as
\begin{equation}
n=\kappa h^{\frac{1}{\gamma-1}},\label{density}
\end{equation}

\noindent where $\kappa=\left(\frac{\gamma-1}{\gamma 
K}\right)^{\frac{1}{\gamma-1}}$. In this case, equation (\ref{diffeqn}) 
can be written as
\begin{equation}
\Box \Phi + {\frac{2 - \gamma}{\gamma -1}} [\ln h(\Phi)]_{,\alpha}
\Phi^{,\alpha}=0. \label{diffnlin}
\end{equation}

The simplest case is found when $\gamma = 2$, Eq.~(\ref{diffeqn}) 
reduces to a linear equation. In this case the barotropic sound speed, 
defined as ${\rm d}p/{\rm d}\rho \equiv c_s^2$, is equal to the speed of 
light, i.e., we have a stiff equation of state.  In the general case 
($1<\gamma <2$), the differential equation is nonlinear and the 
barotropic sound speed, $c_s^2=(\gamma -1)$, is less than the speed of 
light.

The normalization condition, $U_\alpha U^\alpha =-1$, gives us a 
relation between the enthalpy and the scalar field,
\begin{equation}
h=\sqrt{-\Phi{,_\alpha} \Phi^{,\alpha}}. \label{enthalpy}
\end{equation}

\noindent This relation will be also useful to determine the baryon 
number density.
 
In this section we have followed the work of Moncrief \cite{mon} which 
is in accord with Tabensky and Taub \cite{tab:tau} who consider a 
constant barotropic sound speed, $c_s^2=(\gamma -1)$.

\section{Potential flows in Reissner-Nordstr\"om-(anti) de Sitter black 
holes and rigid spheres}

\subsection{The metric}

The Reissner-Norstr\"om-(anti) de Sitter metric can be written as
\begin{equation}
ds^2=-f{\rm d}t^2+f^{-1}{\rm d}r^2+r^2{\rm d}\Omega^2,\label{metric}
\end{equation}

\noindent where $f=1-2M/r + Q^2/r^2 - \Lambda r^2/3$, $M$ is the mass, 
$Q$ is the electric charge and $\Lambda$ is the cosmological constant. 
The metric (\ref{metric}) has different number of horizons given by the 
roots of $f=0$, which depend on $M$, $Q$ and $\Lambda$. If $\Lambda>0$ 
(de Sitter spacetime) we may obtain four different real roots. For 
simplicity, we have taken our parameters such this is always true. Three 
of the roots are then identified as the inner (or Cauchy) horizon 
($r_i$), the event (or outer) horizon ($r_h$) and the cosmological 
horizon ($r_c$). These horizons satisfy the relation $r_i<r_h<r_c$. The 
remaining root is negative and no physical meaning is attached. 
Following the classification due to Brill and Hayward \cite{brill}, we 
are dealing with a generic black hole. If $\Lambda<0$ (anti de Sitter 
spacetime), we obtain two real roots representing the inner horizon and 
the event horizon. In this case there is no cosmological horizon and the 
function $f(r)$ diverges as the radial coordinate $r$ tends to infinity.

\subsection{Numerical method}

To solve equation (\ref{diffnlin}) in the space-time with metric 
(\ref{metric}) we assume: (a) That fluid is stationary, i.e., the 
function $\Phi$ depends on time only through the addition of $-at$, 
where $a$ is a constant related to the zeroth component of the velocity. 
(b) That due to the axial symmetry of the metric the potential $\Phi$ 
does not depend on the variable $\varphi$, and (c) That the fluid is a 
test fluid, i.e. the metric does not evolve and it is given {\em a 
priori}. With the above mentioned assumptions Eq.~(\ref{diffnlin}) 
reduces to an elliptical differential equation with an inner boundary 
condition near the black hole or rigid sphere and an external boundary 
condition (asymptotic condition).

To obtain the numerical solution of Eq.~(\ref{diffnlin}) we use a 
computational code based on a relaxation method with a second-order 
precision, five-point, finite difference scheme. The relaxation method 
employed is the \textit{Stone's strongly implicit procedure} applied in 
a numeric grid evenly spaced in both $r$ and $\theta$. The whole method 
is detailed in reference \cite{ferziger}.

In the case of a rigid sphere, for the inner boundary condition, the 
usual condition of zero normal velocity in the surface of the sphere is 
employed.  In general, the fluid velocity must be equal to the 
corresponding component of the velocity of the surface.  Since usual 
stars have gaseous surfaces (not hard), this condition does not describe 
a typical flow around a star.  In special astrophysical situations, like 
relativistic flows around a neutron star, this condition can be valid.  
For a black hole we use the condition that the fluid particle number 
density must remain finite on the black hole horizon \cite{pet:sha}. 
This leads to the numerical condition \cite{abr:sha} that near $r=r_h$ 
(the black hole horizon),
\begin{equation}
\frac{\partial}{\partial r^*} \left[ \frac{\Phi_{,r^*}
- \Phi_{,t}}{r-r_h} \right] =0, \label{condtort}
\end{equation}

\noindent where ${\rm d}r^*\equiv {\rm d}r/f$ defines the tortoise 
radial coordinate \cite{reg:whe}.  With the assumption that 
$\Phi_{,t}=-a<0$, the condition (\ref{condtort}) can also be satisfied 
in our case and it will be taken as the inner boundary condition. For 
numerical applications of this condition see \cite{uje:let1,fon:mar}.

On the outer boundary we apply the dirichlet condition imposing that the 
fluid is homogeneous and has a constant velocity parallel to the z-axis 
of the coordinate system, say
\begin{equation}
\Phi=-U_\infty^0t+U_\infty r\cos\theta,
\end{equation}

\noindent where $U^a_\infty=(U^0_\infty,\mathbf{U_\infty})= 
(1-v_\infty^2)^{-1/2}(1,\mathbf{v_\infty})$. This condition is applied 
in a region located usually between the surface of the sphere (or black 
hole horizon) and 60 Schwarzschild radius. Note that the presence of a 
cosmological horizon (for $\Lambda >0$) does not represent any further 
difficulty. Our domain of interest is delimited to the asymptotic 
boundary condition, which we carefully apply in a radius much smaller 
than $r_c$.

To solve the non-linear equation (\ref{diffnlin}) we first put the 
non-linear term equal to zero and calculate the linear part of the 
equation to find $\Phi$. This solution is used as an initial guess for 
the non-linear problem. Then, from (\ref{enthalpy}) we compute the fluid 
enthalpy.  Finally, with this information we compute the non-linear term 
that is introduced in the non-linear equation to find new values of 
$\Phi$. The process is repeated until the sum of the fractional change 
in the enthalpy for all interior points of the grid in one iteration is 
less than a certain error $\xi$. Usually less than 10 iterations were 
required to reach an error of $\xi \leq 10^{-6}$.  Obviously, for the 
linear case we need only one iteration. The number of iterations depends 
on the value of $\gamma$. We found that for values of $\gamma < 5/3$ our 
code does not converge, this is because the enthalpy becomes negative in 
some points of our grid. So, this value of $\gamma$ may represent a 
physical limit to our model. Moreover, when $\gamma=1$ the fluid is 
pressureless (dust) and the fluid flow is geodesic, i.e., no longer 
obeys (\ref{diffnlin}).

The code was tested for the case $\gamma=2$ using the analytical results 
for the steady flow of a fluid in the presence of either a hard sphere 
\cite{sha} or a black hole \cite{pet:sha,bab:che}. In this case the 
numerical solution agreed with the exact solution within an error better 
than 1\% for the radial velocity and angular velocity. This accuracy is 
sufficient for our purposes.

Another aspect of the fluid dynamics is the accretion rate of matter 
into a black hole.  This accretion rate can be computed from 
(\ref{potflow}) and (\ref{density}), we find
\begin{equation}
\dot{N} = -\int_S nU^i \sqrt{-g} dS_i = - \int_S \kappa h^{ 
\frac{2-\gamma}{\gamma-1}}\Phi_{,r} g^{rr} \sqrt{-g} d\Omega,
\label{accretion}
\end{equation}

\noindent where the integration is performed in a two-surface sphere 
centered on the black hole.  The exact form of (\ref{accretion}), for 
$\gamma=2$, is known \cite{pet:sha,bab:che}. It is $\dot{N}=4\pi r_h^2 
n_\infty U_\infty^0$, where $r_h$ is the event horizon and $n_\infty$ 
and $U_\infty^0$ are the asymptotic density and zeroth component of the 
4-velocity, respectively. The quantity $n_\infty$ is a constant in this 
case. We test our code to compute the accretion rate with this exact 
solution. We found an error less than 1\%.

\begin{figure*}
\hspace{-0.3cm}
\epsfig{scale=0.67, file=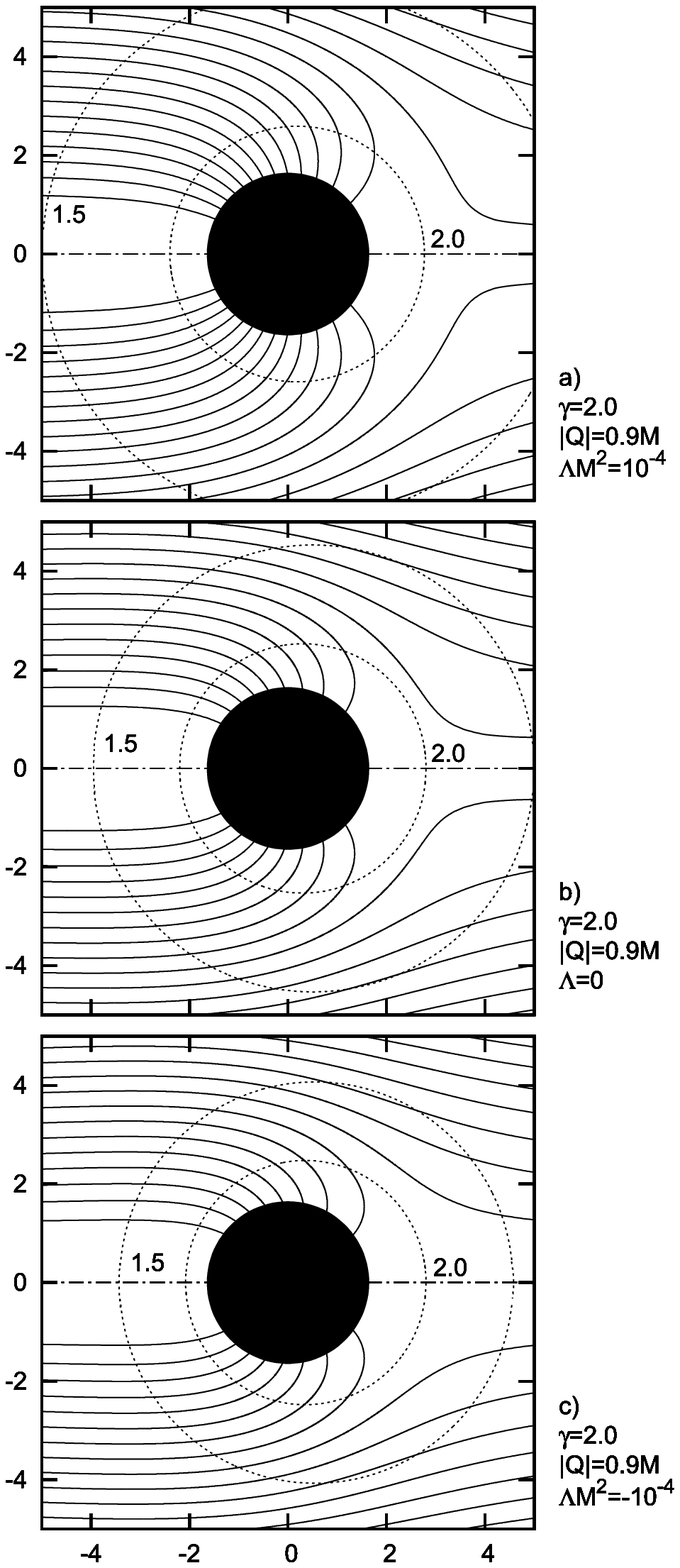} 
\hspace{0.1cm}
\epsfig{scale=0.67, file=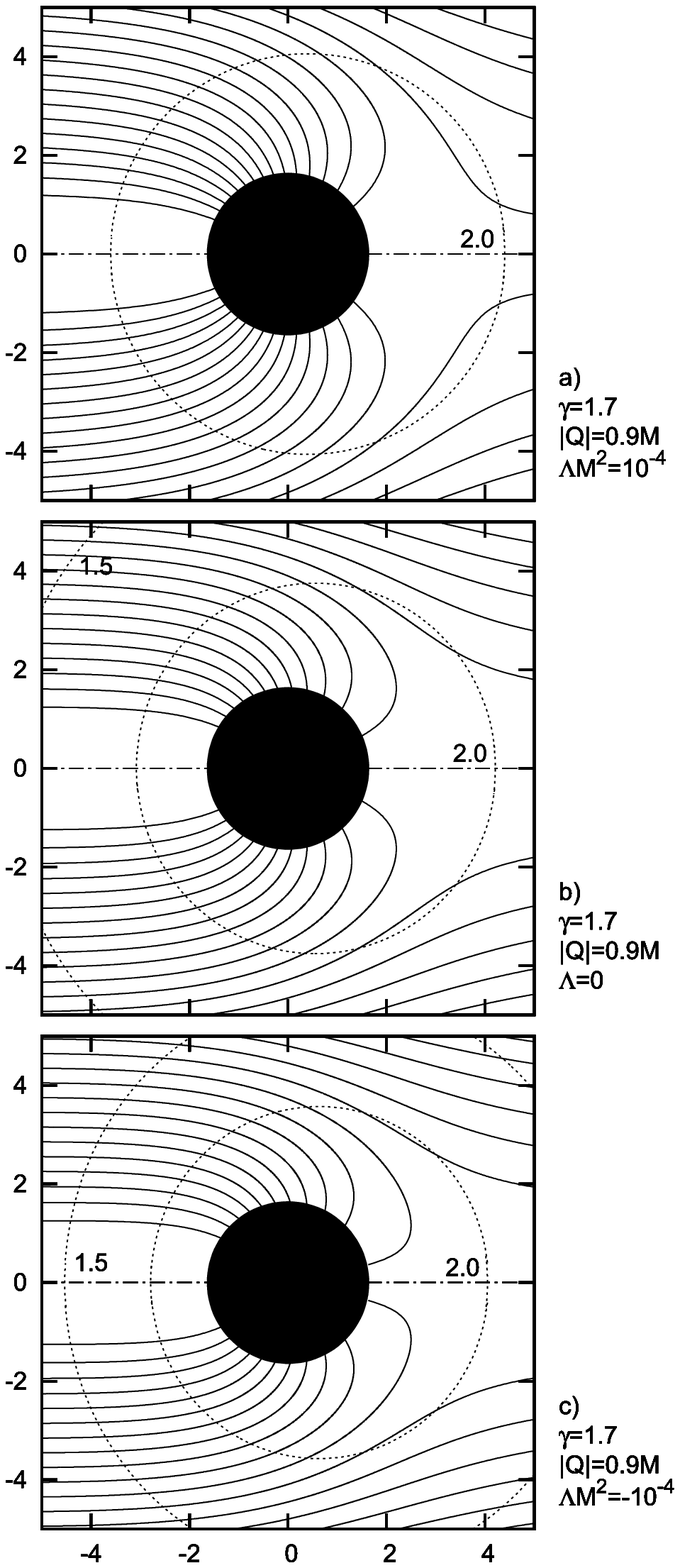}

\caption{Numerical results for the streamlines and density contours 
$(n/\kappa)$ for a stiff fluid ($\gamma=2$) and a fluid with 
$\gamma=1.7$ flowing through 
a black hole with an electric charge $Q=0.9M$. Three values for 
$\Lambda M^2$ were considered for comparison (null, positive and 
negative cosmological constant). The fluid flows from left to right 
with $v_\infty=0.6$.}
\label{fig1} 
\end{figure*}

\begin{figure*}
\hspace{-0.3cm}
\epsfig{scale=0.67, file=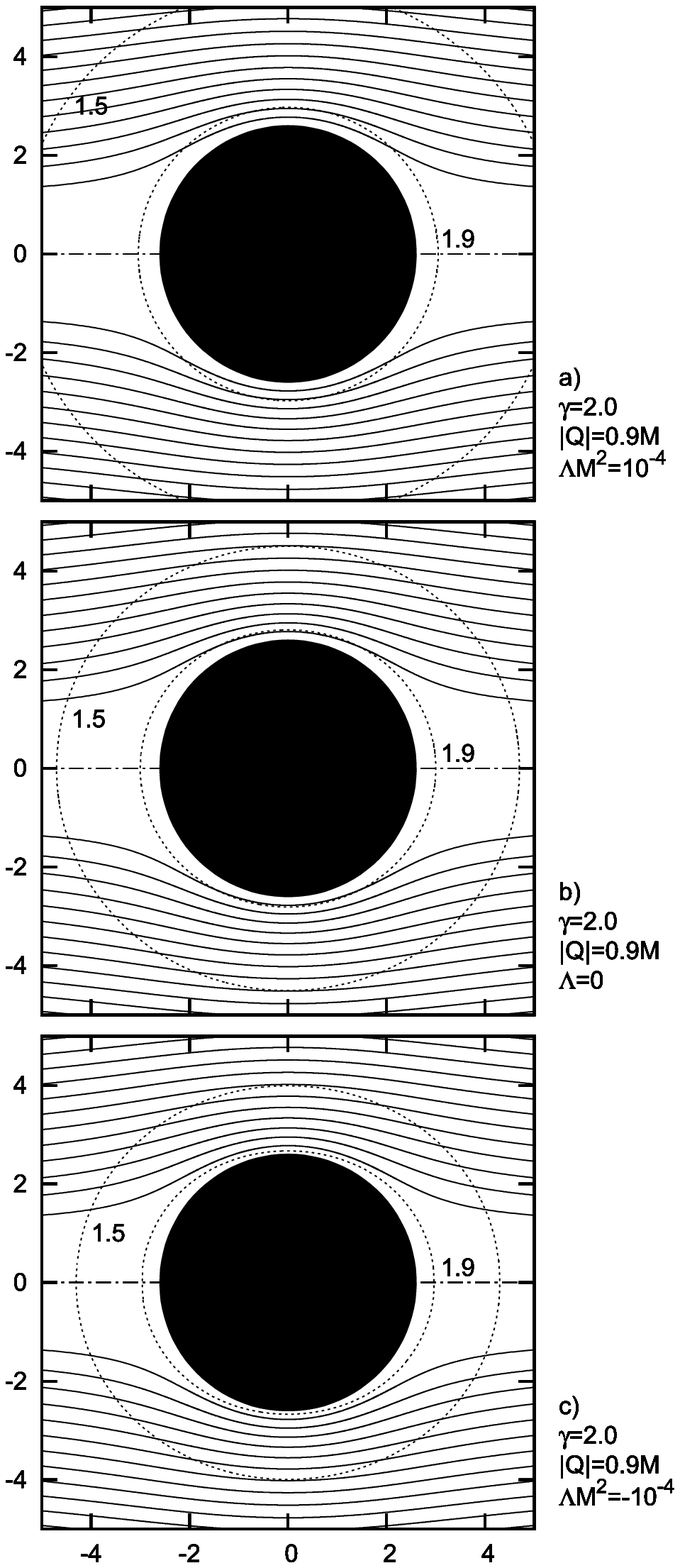} 
\hspace{0.1cm}
\epsfig{scale=0.67, file=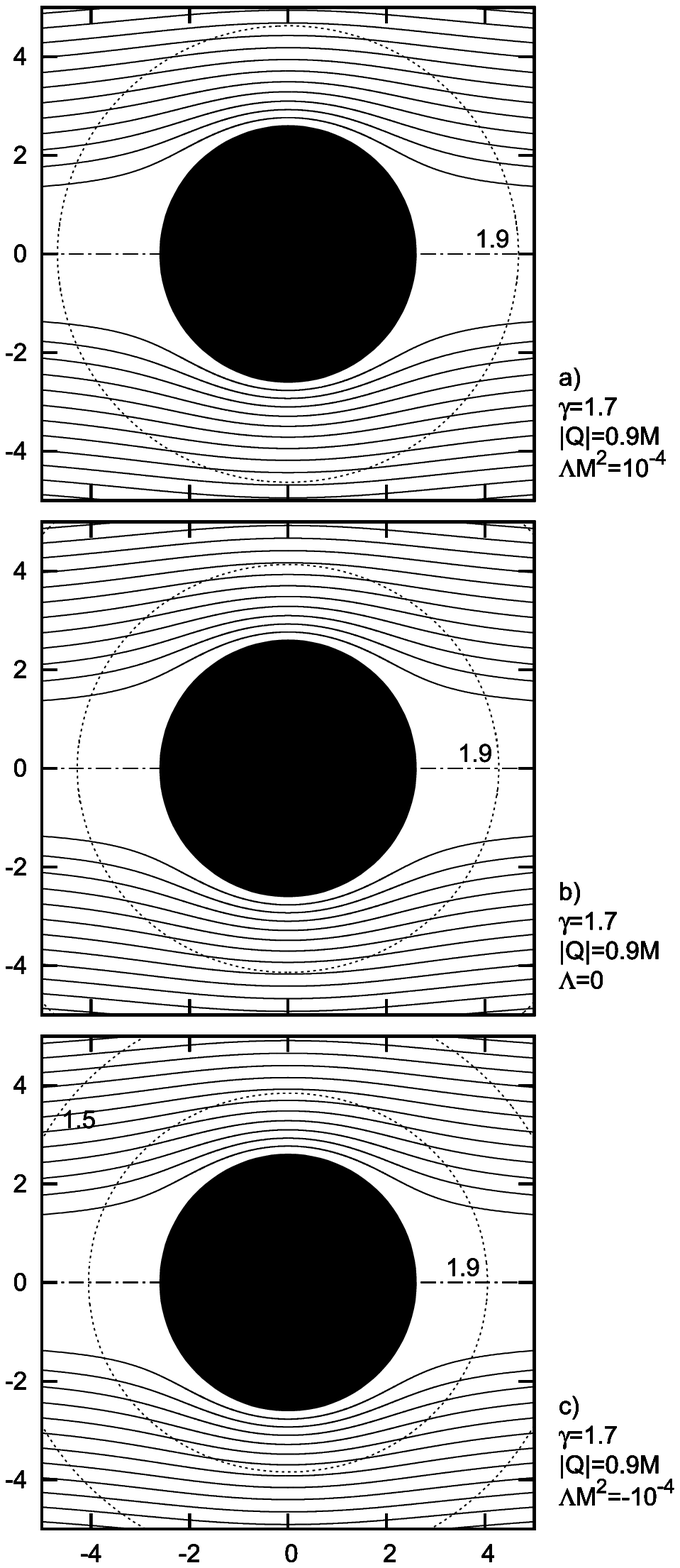}
\caption{Numerical results for the streamlines and density contours 
$(n/\kappa)$ for a stiff fluid ($\gamma=2$) and a fluid with 
$\gamma=1.7$ flowing through a hard sphere of radius $R=2.5M$ with an 
electric charge $Q=0.9M$. Two values are considered for $\Lambda M^2$ to 
contrast to the case of $\Lambda=0$. The fluid flows from left to right 
with $v_\infty=0.6$.}
\label{fig2} 
\end{figure*}

\begin{table*}
\begin{minipage}{\textwidth}
\caption{Qualitative description of the influence of the parameters 
$(Q,\Lambda,\gamma)$ and their combined changes on the potential flow
through black holes and rigid spheres.}
\begin{ruledtabular} \label{sum_table}
\begin{tabular}{p{1.8cm}p{3.3cm}p{5.6cm}p{5.6cm}}
\textbf{Variable parameter} & \textbf{Variation performed} & 
\textbf{Summary of effects:               Black holes} 
                                                           & 
\textbf{Summary of effects: Rigid spheres} \\
\hline
& & &\\
$Q$\footnotemark[1] & Increase of the ratio $Q^2/M^2$, with $\Lambda=0$.
                    & Decrease of $\dot{N}$. 
                      Increase of the density around the black hole. 
                      Expansion of the density contours. 
                    & Decrease of the density around the sphere. 
                      Contraction of the density contours.\\
& & &\\
$\Lambda$ & Increase/Decrease of $\Lambda M^2$ and $Q=0$.
          & Increase/Decrease of $\dot{N}$. Increase/Decrease of the 
            density over the region of 
            incidence 
            of the fluid and decrease/increase of the density on the 
            opposite region. Slight expansion/contraction and shift to 
            the left/right of the density contours. Increase/Decrease 
            of the number of streamlines that fall into the black hole.
          & Increase/Decrease of the density around the sphere, mainly 
            over the region of $\theta=\pi/2$.
            Expansion/Contraction of the density contours. \\
& & &\\
$Q,\Lambda$ & Increase/Decrease of $\Lambda M^2$ and $Q\neq 0$.
            & The effects due to the presence of $\Lambda$
              are qualitative the same as those observed if $Q=0$ but
              less evident.
            & The effects due to the presence of $\Lambda$
              are qualitative the same as those observed if $Q=0$ but
              more evident.\\
& & &\\
$\gamma$\footnotemark[2] & Decrease of $\gamma$ to values smaller than 
$2$.
                         & Increase of $\dot{N}$.
                           Increase of the density around the black 
                           hole. 
                           Expansion of the density contours.
                           Increase the number of streamlines that 
                           fall into the black hole. The effects due to 
                           variations of $Q$ and/or $\Lambda$ are 
                           enhanced.
                         & Increase of the density around the sphere.
                           Expansion of the density contours. The 
                           effects due to variations of $Q$ and/or 
                           $\Lambda$ are enhanced.
\end{tabular}
\end{ruledtabular}
\footnotetext[1]{This case has analytical solutions for $\gamma=2$ 
\cite{bab:che}. }
\footnotetext[2]{The effects of $\gamma<2$ had been treated 
in other physical situations. See for example
references \cite{abr:sha,uje:let2}}
\end{minipage}
\end{table*}

\section{Numerical results}

\subsection{Black holes}

In Fig.~\ref{fig1}, we show the streamlines and density contours 
($n/\kappa$) for a potential flow with $\gamma =2$ and $\gamma =1.7$ 
through a Reissner-Nordstr\"om black hole for different values of the 
factor $\Lambda M^2$. For increasing values of $\Lambda M^2$, and fixed 
$Q$ and $\gamma$, we see, from the upper and middle graphs of 
Fig~\ref{fig1}, that the density in the incident region increases and in 
the opposite side decreases. On the other hand, for decreasing values of 
$\Lambda M^2$, we obtain, from the middle and lower graphs of 
Fig.~\ref{fig1}, that the density decreases in the incident region and 
increases in the opposite side. These behaviors can be seen as a shift 
of the density contours (more evident for $n/\kappa=2.0$) and an 
expansion or contraction of the density contours (more evident for 
$n/\kappa=1.5$). These effects are enhanced for lower values of 
$\gamma$, e.g. the contour $n/\kappa=1.5$ does not even appear in the 
upper graph when $\gamma=1.7$. It must be noted, however, that the 
adopted values for $\Lambda M^2$ are choose in order to magnify their 
effects on the flow.

In Fig.~\ref{fig1} we use a fixed value for the electric charge, say 
$Q=0.9M$. Further simulations show that the qualitative properties of 
the density contours are independent whether the black hole has electric 
charge or not. Nevertheless, the density contour values are more 
sensitive to changes in $\Lambda$ if the black hole has small or null 
electric charge. In general, all the effects are more evident for lower 
values of $\gamma$.

Another quantity of interest for flows through a black hole is the 
accretion rate of matter given by Eq.~(\ref{accretion}). We numerically 
integrate Eq.~(\ref{accretion}), using Simpson's rule, to find that the 
presence of a positive or negative cosmological constant leads to an 
increase or decrease in the accretion rate, respectively. Furthermore, 
the increase due to $\Lambda>0$ is numerically equivalent to the 
decrease due to $\Lambda<0$. The accretion rate is slightly larger if 
the black hole has small or null electric charge.

Regardless the value of $\Lambda$ and $Q$, the accretion rate always 
increases for lower values of $\gamma$. Also the accretion rate is more 
sensitive to variations in both $Q$ and $\Lambda$ if $\gamma$ is less 
than 2. For example, if $Q=0.9M$, the influence of the cosmological 
constant is about ten times greater when we lower the value of $\gamma$ 
from 2 to 1.7, and for $Q=0$ is almost twenty times greater. Moreover, 
examining the streamlines pattern, we note that, if we increase or 
decrease the value of $\Lambda$ we also increase or decrease the number 
of streamlines that fall into the black hole, respectively, see 
Fig~\ref{fig1}. For lower values of the parameter $\gamma$ the 
streamlines flux through the black hole increases, this behavior is in 
accordance with the previous numerical results for the accretion rate.

\subsection{Rigid spheres}

In Fig.~\ref{fig2}, we show the streamlines and density contours 
($n/\kappa$) for a potential flow with $\gamma =2$ and $\gamma =1.7$ 
through a Reissner-Nordstr\"om rigid sphere for different values of the 
factor $\Lambda M^2$. We see that for increasing values of $\Lambda 
M^2$, with fixed $Q$ and $\gamma$, the density contours expand, which is 
associated to an increase in the density around the rigid sphere. Also, 
the effect of the cosmological constant is to change the oblateness of 
the density contours. For large values of $\Lambda$ the contours are 
more spherical, this is easily noted from the $n/\kappa=1.9$ density 
contours of the $\gamma=2$ column graphs. When $\Lambda M^2$ decreases, 
we obtain the opposite behavior. These expansions and contractions of 
the density contours are enhanced when we lower the value of the 
parameter $\gamma$, this can be seen through a direct comparison from 
the graphs of the different columns of Fig~\ref{fig2}. For all 
parameters, the streamlines suffer little changes.

In addition, further numerical simulations show us that, for a fixed 
$\gamma$, the density of fluid around a hard sphere is more sensitive to 
changes in $\Lambda$ if the rigid sphere has a large electric charge. 
This is contrary to what happens in the black hole case where we find 
that the effects are less sensitive when it has a large electric charge.

\section{Conclusions}

We have numerically analyzed the effect of the electric charge and both, 
positive and negative, cosmological constants on a potential flow of an 
ideal fluid with polytropic equation of state $p=Kn^\gamma$ around a 
black hole and a hard sphere by means of the streamlines, the fluid 
density contours and the accretion rate (black hole case only). The main 
effects observed in our numerical simulations are summarized in 
Table~\ref{sum_table}.

As a final comment, we must say that results in cosmology indicates with 
99\% of confident that $\Lambda > 0$ \cite{per:etal}, with possible 
values of $\Lambda \approx 10^{-52}$/m$^2$ 
\cite{per:etal,per:etal2,car:kuz}. A value such as $\Lambda M^2 \approx 
10^{-7}$ implies in an object with $M \approx 10^{19}$ solar masses. If 
we considered that the supposed super massive black holes in the center 
of some galaxies have masses of order $M \approx 10^{-9}$, we conclude 
that the detection of the influence of the cosmological constant near 
such objects is going to be difficult.

\section*{Acknowledgments}

We want to thank UFABC and CAPES for financial support.

\end{document}